%% file: frascrpp.tex
\documentstyle[twoside,epsfig,fleqn,espcrc2]{article}

\newcommand{\beq}{\begin{equation}}
\newcommand{\eeq}{\end{equation}}
\newcommand{\bea}{\begin{eqnarray}}
\newcommand{\eea}{\end{eqnarray}}

  \input tables

\newcommand{\AmS}{{\protect\the\textfont2
  A\kern-.1667em\lower.5ex\hbox{M}\kern-.125emS}}

\title{Cosmic Ray particle production
      }

\author{J.~Ranft\address{INFN, Lab. Naz. del Gran Sasso, 
I--67010 Assergi, Italy
        }%
        \thanks{Present adress: FIGS and Physics Dept. 
	Universit\"at Siegen, D--57068 Siegen Germany,
	e--mail: 
	Johannes.Ranft@cern.ch
                }
        }
       
\begin{document}

\begin{abstract}
The status of some popular models to simulate hadronic and nuclear interactions at Cosmic Ray energies is reviewed.
 The models predict
the rise of all the hadronic and nuclear cross sections with
energy and a smooth (logarithmic) rise of average multiplicities,
rapidity plateaus and average transverse momenta with the
energy. Big differences are found between  model predictions partly already 
at energies, where collider data are available. It is argued, that 
at the highest energies data of the Cosmic Ray cascade can 
only be reliably interpreted by sampling the cascade using more than one model.
The importance is stressed to put more effort into the models and 
especially a better understanding of the minijet component at the 
highest energies. Likewise, experimental data on particle production
are needed at the highest possible energies, to guide the models. 
 
{\it Presented at the International Symposium om Multiparticle
Dynamics Frascati, Italy Sept.8 to 12 1997} 

\end{abstract}

\maketitle

\section{Introduction}

The extension of models for multiparticle production in hadron--hadron,
 hadron--nucleus and nucleus--nucleus collisions to be used for 
the simulation of the Cosmic Ray cascade up to $E_{lab}$ = 10${}^{21}$ eV
(corresponding to $\sqrt s $ = 2000 TeV) is needed to prepare for the 
Auger experiment \cite{Auger} as well as for a reliable interpretation 
of present experiments like Agasa \cite{Agasa} and Flys Eye \cite{Flyseye}, 
which present data in the EeV energy region. The need for careful comparisons
of hadron production models was stressed at the International Cosmic Ray 
Conference in Roma 1995. Following this, such a code comparison in the 
energy region of interest to the Kaskade experiment \cite{Kaskade} was 
presented by members of the Kaskade experiment \cite{Kcodcomp}. 
From this code comparison it became clear, that already in the knee region 
of the Cosmic Ray energy spectrum important differences exist between
the models and that these differences might change the interpretation of
certain Cosmic Ray results.
Here we will discuss the status of some of these models, discuss
the minijet component, present typical comparisons to Collider
data, present some characteristics of hadron production up to
$E_{lab}$ = 10${}^{21}$ eV 
and finally compare some results obtained simulating the
cosmic ray cascade using different models.

\section{The present status of some event genera\-tors used 
for Cosmic Ray cascade simulations}

The presently dominant hadron production models used for the
simulation of the Cosmic Ray cascade are constructed on the
basis of multistring fragmentation, they use Gribov--Regge and
Gribov--Glauber theory, to construct the multistring production
in hadron--hadron and nuclear collisions. Most of the models use
minijets as an important mechanism for particle production at
high energies.

 The DPMJET--II
event generator based on the two--component Dual Parton Model
(DPM) was described in detail 
 \cite{DPMJETII,Ranftsare95,Dpmjet23}.
 The extension of this model up to energies of $\sqrt s$ =
2000 TeV was reported this year, 
the resulting model  will be refered to as DPMJET--II.3.
The extension is done by calculating the minijet component of
the model using new parton distribution functions, the
 GRV--LO parton distributions
\cite{Gluck95a}  
and the CTEQ4 parton distributions \cite{Lai97},
which are both available in a larger Bjorken--$x$ range than the
MRS(D-) parton distributions, which were the default in
DPMJET--II.2. These new parton
distributions describe the structure function 
data measured in the last years at the
HERA Collider.
  DPMJET--II.3 
 descibes well minimum bias hadron and hadron jet
 production up to  present collider energies. It is also
 demonstrated, that the model performs as well as the
 previous one DPMJET--II.2  for hadron production in
 hadron--nucleus and nucleus--nucleus colisions.
 DPMJET is used for the simulation of the Cosmic Ray cascade
 within the HEMAS--DPM code \cite{DPMBFR94} used mainly for the
 MACRO experiment \cite{Macro}.

 The SIBYLL model \cite{SIBYLL} is a minijet model and has been
 reported to be applicable up to $E_{lab}$ = 10${}^{20}$ eV.
 However, the EHQL \cite{EHQL} parton structure functions used
 for the calculation of the minijet component might , after the
 HERA experiments, no longer be adequate. It is known, that a
 significant updating of SIBYLL is planned for the next year.
 SIBYLL is the most popular model for simulating the Cosmic Ray
 cascade in the USA.

 VENUS, a very popular model applied originally for describing
 heavy ion experiments, is now the leading event generator
 within the Corsika Cosmic Ray cascade code \cite{Corsika}.
 VENUS is appicable up to $E_{lab}$ = 5$\times10{}^{16}$ eV.
 It has been reported \cite{Werner96}, that the introduction of
 minijets into VENUS has been planned, this will allow to apply
 VENUS up to higher energies.

 QGSJET \cite{QGSJET} is the most popular Russian event
 generator used for Cosmic Ray simulations. It is based on the
 Quark Gluon String (QGS) model, this model is largely
 equivalent to the DPM. QGSJET also contains a minijet component
 and is reported to be applicable up to $E_{lab}$ = 10${}^{20}$
 eV.

 HPDM \cite{HPDM} is based on parametrizations inspired by the
 DUAL Parton Model. it is reported to be applicable up to
 $E_{lab}$ = 10${}^{20}$ eV, howeversome of the parametrizations
 might become unreliable above $E_{lab}$ = 10${}^{17}$ eV. HPDM
 was originally used as event generator within the Corsika
 cascade code.

MOCCA \cite{MOCCA} is an empirical model employing a succesive
splitting algorihm. It was reported to be applicable up to
$E_{lab}$ = 10${}^{20}$ eV. Since the model does not contain
minijets, its predictions at the upper energy end might differ
significantly from the other models.

In Table 1 we present some characteristics of the models. The
Gribov--Regge theory is applied by three of the models. The
pomeron intercept for SIBYLL is equal to one, SIBYLL is a
minijet model using a critical pomeron, with one soft chain
pair, all the rise of the cross section results from the
minijets. In the models with pomeron intercept bigger than one,
we have also multiple soft chain pairs, already the soft pomeron
leads to some rise of the cross sections with energy. Minijets
are used in three of the models, it is believed, that minijets
are
necessary to reach the highest energies. All models contain
diffractive events. Secondary interactions between all produced
hadrons and spectators exist only in VENUS, DPMJET has only a
formation zone intranuclear cascade (FZIC) between the produced
hadrons and the spectators. Only three of the models sample
properly nucleus--nucleus collisions, the other two models
replace this by the superposition model, where the
nucleus--nucleus collision is replaced by some
hadron--nucleus collisions. The residual projectile (and target)
nuclei are only given by two of the models.

{\bf Table 1.}
Characteristics of some popular mo\-dels for hadron production  in
Cosmic Ray cascades. (VEN = VENUS, QGS = QGSJET, SIB = SIBYLL, HP =
HPDM, DPM = DPMJET) 
\vskip 5mm
\begintable
 {}           |VEN | QGS | SIB | HP | DPM   \cr
 Grib.--Regg. | x  | x   |     |    |  x    \cr
 Pom. ic.     |1.07|1.07 | 1.00|    | 1.05  \cr
 minijets     |    |  x  |  x  |    |  x    \cr
 Diffr. ev.   | x  |  x  |  x  | x  |  x    \cr
 sec. int.    | x  |     |     |    |  x    \cr
 A--A int.    | x  |  x  |     |    |  x    \cr
 superp.      |    |     |  x  | x  |       \cr
 res. nucl.   |    |  x  |     |    |  x    \cr
 max. E [GeV] |$10^7$ | $10^{11}$ |$10^{11}$ | $10^8$ |$10^{12}$  \endtable

\section{The calculation of the minijet component
}
The input cross section (before the unitarization procedure
applied by the models) for semihard multiparticle
production (or minijet production)  $\sigma_{h}$ is calculated applying
the QCD improved parton model,
the details (for DPMJET) are given   
in Ref.\cite{CTK87,DTUJETZP91,DTUJETPR92,DTUJET92b,Hahn90,DTUJET93}.
\begin{eqnarray}
\sigma_h &=& \sum_{i,j}
\int_0^1 dx_1 \int_0^1
dx_2 \int d\hat{t}\ 
 \frac{1}{1+\delta_{ij}} 
\frac{d\sigma_{QCD,ij}}{d\hat{t}}\ 
\nonumber\\
& &\times
f_i(x_1,Q^2)
f_j(x_2,Q^2)\ \Theta(p_\perp-p_{\perp{thr}})
\end{eqnarray}
$f_i(x,Q^2)$ are the structure functions of partons with the flavor
$i$ and scale $Q^2$ and the sum $i,j$ runs over all possible flavors.
To remain in the region where perturbation theory is valid, 
 a low $p_{\perp}$ cut--off $p_{\perp_{thr}}$  is used 
for the minijet component.
Since the HERA measurements, the structure functions are known
to behave at small $x$ like 1/$x^{\alpha}$ with $\alpha$ between
1.35 and 1.5.
The minijet production is dominated by very small $x$ values,
therefore the minijet cross section calculated with the new
structure functions rise very steeply with energy. We found
already 1993 \cite{DTUJET93} with the MRS[D-] structure function
\cite{MRS92} at the LHC energy a minijetcross section about 10
times larger than the total cross section at this energy. 
\begin{figure}[thb]
\begin{center}
\epsfig{file=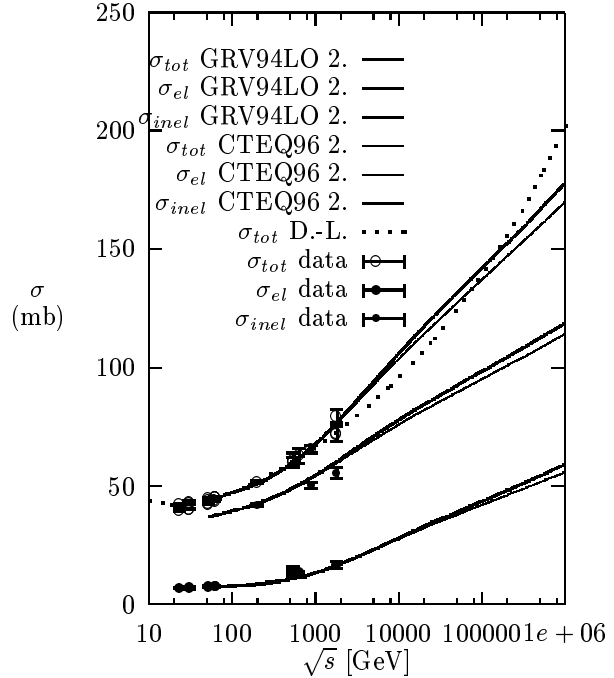,width=8cm,height=9cm}
\end{center}
\vspace*{-3mm}
\caption{Total, inelastic and elastic $p\bar p$ and $pp$ cross
sections from DPMJET--II.3 
as function of the center of mass energy $\sqrt s$. The
model results obtained using the GRV--LO parton distributions
\protect\cite{Gluck95a} and the CTEQ4 parton distributions 
\protect\cite{Lai97}
are compared to the Donnachie--Landshoff fit for the total cross
section \protect\cite{Donnachie93} and to data 
\protect\cite{Arnison83,Bozzo84a,Amos85}
\protect\cite{Bernard87a,Alner86a,Amos90a,Abe93a,Abe94b,Abe94d}
\protect\label{xs222626}
}
\end{figure}
\begin{figure}[thb]
\begin{center}
\epsfig{file=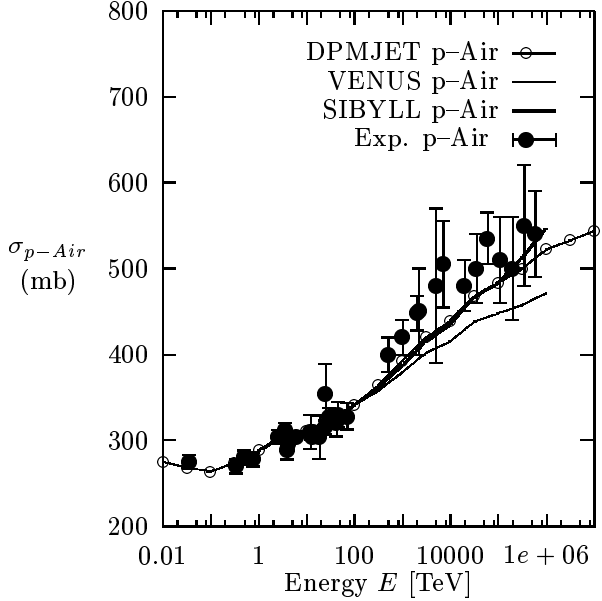,width=8cm,height=8cm}
\end{center}
\vspace*{-3mm}
\caption{The inelastic cross section $\sigma _{p-Air}$ calculated
by DPMJET--II.3 (as well as the ones from VENUS and SIBYLL
according to Ref. \cite{Kcodcomp}) as
function of the laboratory collision energy (from 0.02 TeV up to
1.E9 TeV) compared to
experimental data collected by 
Mielke et al. \protect\cite{Mielke}. 
\protect\label{sigpair}
}
\end{figure}
Such large minijet cross sections are inconsistent and wrong:
The input minijet cross sections $\sigma_h$, 
which one puts  into the unitarization scheme
are 
inclusive cross sections normalized to $n_{minijets}\sigma_{inel}$,
where $n_{minijets}$ is the multiplicity of minijets.
 The physical processes, which contribute to this
inclusive cross section
 are $2 \rightarrow n$ parton processes.
$2 \rightarrow n$ processes
give a contribution to $\sigma_h$ equal to $n\sigma_{2 \rightarrow n}$.
If one treats this huge cross section 
as $\sigma_h$ in the usual way in the
 eikonal unitarization scheme one replaces it by $ n/2$
 simultaneuos $2 \rightarrow 2$ parton processes, 
  this is the
 inconsistency. 
What one should really use in the unitarization, but what
we do not know how to compute reliably at present would be
$\sigma_h = \sum_n \sigma_{2 \rightarrow n}$.
The way to remove this inconsistency is to make in  the two
component DPM
the threshold for minijet
production  $p_{\perp thr}$ energy dependent in such a way, that
at no energy and for no PDF the resulting $\sigma_h$ is 
much 
bigger than
the total cross section. Then at least we have a cross section,
which is indeed mainly the cross section of a $2 \rightarrow 2$
parton process at this level, the parton--parton scattering with
the largest transverse momentum. We can get
back to the real $ 2 \rightarrow n$ processes  and recover the
minijets with smaller transverse momenta via parton showering. One
possible form for this energy dependent cut off 
 is \cite{DTUJET93}:
\begin{eqnarray}
 p_{\perp thr} &=& 2.5 + 0.12[\lg_{10}(\sqrt s/\sqrt s_0)]^3~~~
 \nonumber\\
  & &
 {\rm [GeV/c]},~~~ \sqrt s_0 = 50 {\rm GeV}.
\end{eqnarray}
 The resulting $\sigma_h$ are  smaller  or not much larger than the 
total cross sections resulting after the unitarization 
for all  PDF's.

 \begin{figure}[thb]
 \begin{center}
\epsfig{file=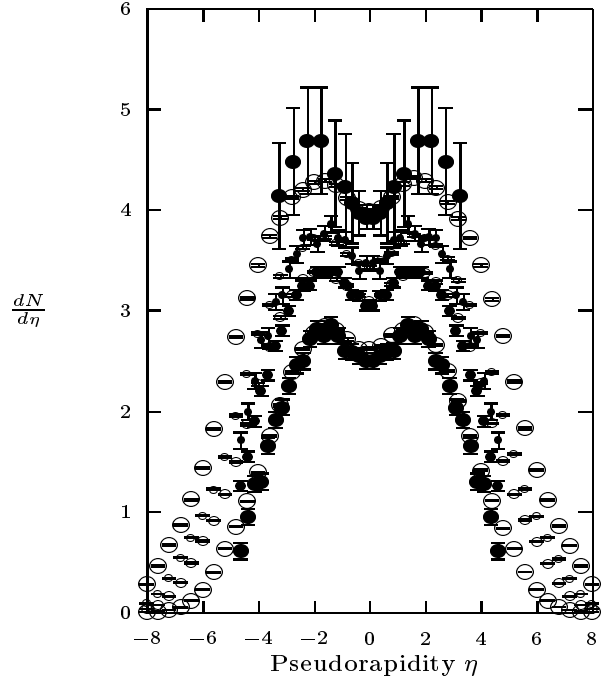,width=8cm,height=9cm}
 \end{center}
 \vspace*{-3mm}
 \caption{Pseudorapidity distributions of charged hadrons
 produced in nondiffractive $p\bar p$ collisions at $\sqrt s$ =
 0.2, 0.54, 0.9 and 1.8 TeV. The DPMJET--II.3 results are
 compared with data from the UA--5 Collaboration \protect\cite{Alner86b}
 and from the CDF Collaboration \protect\cite{Abe90}.
 \protect\label{prapm22225918nd}
 }
 \end{figure}

 There are further features of the minijet component worth
 mentioning.
One uses as first described in \cite{Hahn90}  at $p_{\perp thr}$ the
continuity requirement for the {\it soft}  and
{\it hard} chain end $p_{\perp}$ distributions.   
Physically, this means, that we  use the soft cross
section to cut the singularity in the minijet $p_{\perp}$
distribution. But note, that this cut moves with rising 
collision energy to higher and higher $p_{\perp}$ values. This
procedure has besides cutting the singularity more attractive
features:

{\bf  (i)} The model results (at least as long as we do not
violate the consistency requirement described above) become
somewhat independent from the otherwise arbitrary $p_{\perp}$
cut--off. This was already demonstrated with DTUJET90
\cite{DTUJETPR92} and cut--offs of 2 and 3 GeV/c. 

 {\bf (ii)}The
continuity between soft and semihard physics is emphasized, there
is no basic difference between soft
and semihard chains besides the technical problem, that 
 perturbative QCD allows only to calculate the semihard component. 

{\bf (iii)} With this
continuity in mind  we feel free to call all chain ends, whatever
their origin in the model, minijets, 
as soon as their $p_{\perp}$ exceeds a
certain value, say 2 GeV/c.

\section{Comparing the models  to data at accelerator and
collider energies}
Each model has to determine its free parameters.
  This can be done by a global fit to all available data of
total, elastic, inelastic, and single diffractive cross sections in the
energy range from ISR to collider experiments as well as to the
data on the elastic slopes in this energy range.
Since there are some differences
in the hard parton distribution functions at small
$x$
values resulting in different hard input cross sections we have
to perform separate fits for each set of
parton distribution functions.
After this stage each model predicts the cross sections also
outside the energy range, where data are available.
In Fig. \ref{xs222626}  we plot for DPMJET--II.3 the fitted
cross sections obtained with  two PDF's together with 
the data. Furthermore we compare the total cross sections
obtained with the popular Donnachie--Landshoff fit \cite{Donnachie93}.
For applications in Cosmic Ray cascade simulations we need in
particular the hadron--Air cross section. in Fig.\ref{sigpair}
we compare data according to
Mielke et al. \cite{Mielke} with the cross sections according to
three models. At low energies all models are describing these
data rather well. At high energies we observe however small
differences bet\-ween the models.

\begin{figure}[thb] \centering
\hspace*{0.25cm}
\noindent
\unitlength1mm
\begin{picture}(160,70)
\epsfig{file=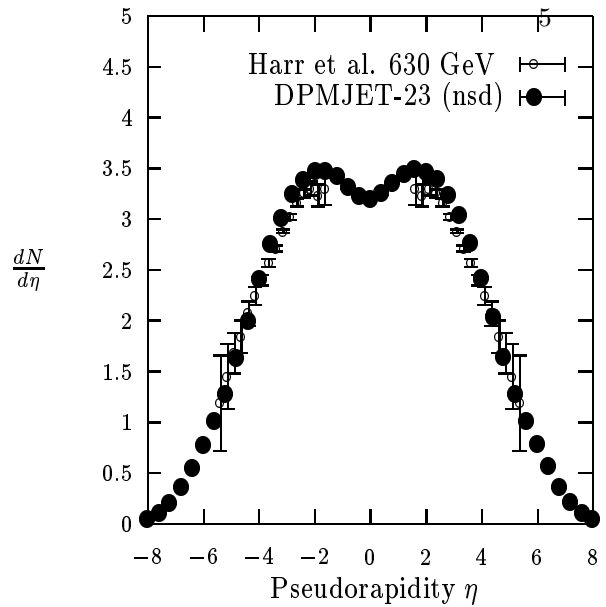,width=8cm,height=8cm}
\end{picture}
\caption{
Pseudorapidity distributions of charged hadrons
produced in nondiffractive $p\bar p$ collisions at $\sqrt s$ =
0.63 TeV. The DPMJET--II.3 results are
compared with  recent data from Harr et al. \protect\cite{Harr97}.
\protect\cite{Adamus88a}.
\label{prapm222630nd}
}
\end{figure}

\begin{figure}
\begin {center}
\epsfig{file=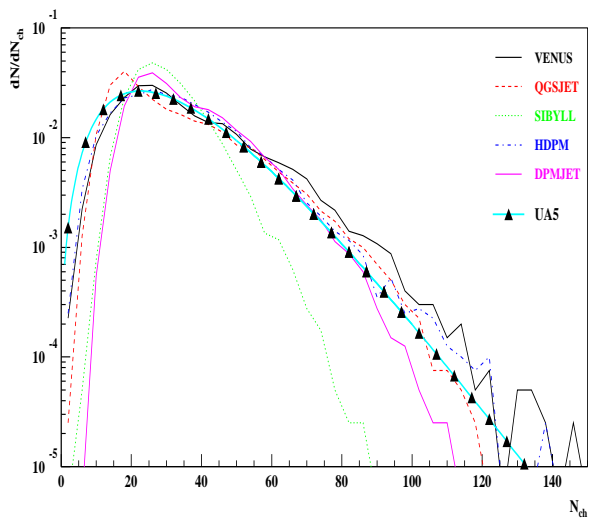,width=8cm,height=7cm}
\end {center}
\vskip -1cm  
\caption{Multiplicity distribution of charged hadrons from
nondiffractive $\bar p - p$ collisions at $\sqrt s $ = 540 GeV. The
data are from the UA--5 Collaboration 
 \protect\cite{Alner86b}. The comparison with 5 models is from 
     \protect\cite{Kcodcomp}.
 }
\label{knapp2}
\end{figure}

 \begin{figure}[thb] \centering
 \hspace*{0.25cm}
 \noindent
 \unitlength1mm
 \begin{picture}(160,70)
\epsfig{file=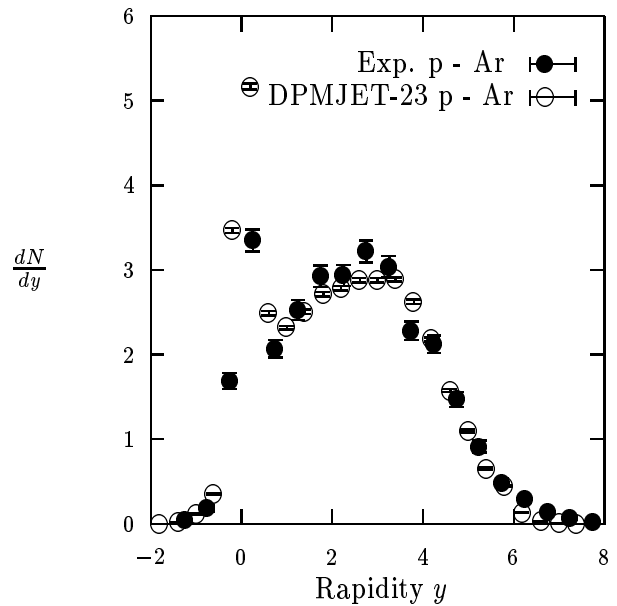,width=8cm,height=8cm}
 \end{picture}
 \caption{
 Charged particle rapidity distribution for p--Ar
 interactions. The DPMJET--II.3 results are compared with data
 \protect\cite{DeMarzo82}.
 \label{rappar}
 }
 \end{figure}

\begin{figure}[thb] \centering
\hspace*{0.25cm}
\noindent
\unitlength1mm
\begin{picture}(80,70)
\epsfig{file=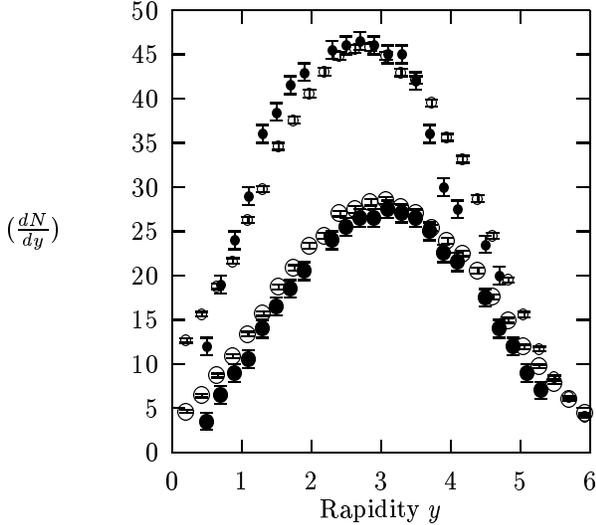,width=8cm,height=7cm}
\end{picture}
\caption{
Rapidity distribution of negatively charged hadrons in
central S--S ans S--Ag collisions. The results of DPMJET--II.3
are compared with data from the NA--35 Collaboration
\protect\cite{na35qm93}.
\label{rapparb}
}
\end{figure}

 \begin{figure}[thb] \centering
 \hspace*{0.25cm}
 \noindent
 \unitlength1mm
 \begin{picture}(160,70)
\epsfig{file=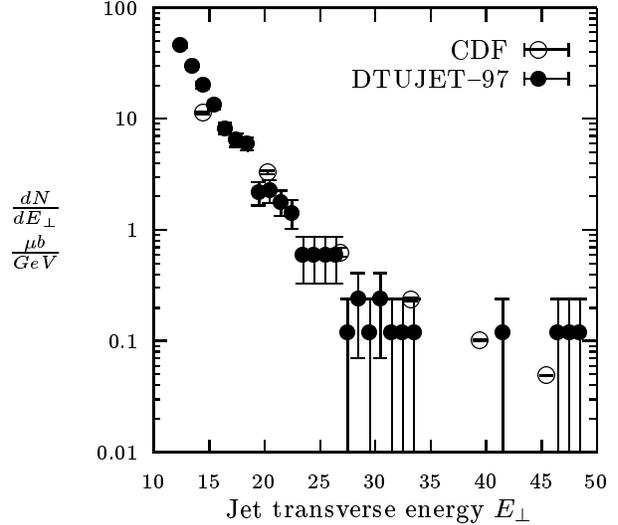,width=8cm,height=7cm}
 \end{picture}
 \caption{
 The jet transverse energy distribution is compared with
 data from the CDF--Collaboration \protect\cite{Abe96a}. The jets
 are found from the model events 
 in the pseudorapidity region $|\eta |\leq$ 0.7  using
 a jet finding algorithm.
 \label{etjet1800}
 }
 \end{figure}

\begin{figure}[thb] \centering
\hspace*{0.25cm}
\noindent
\unitlength1mm
\begin{picture}(160,70)
\epsfig{file=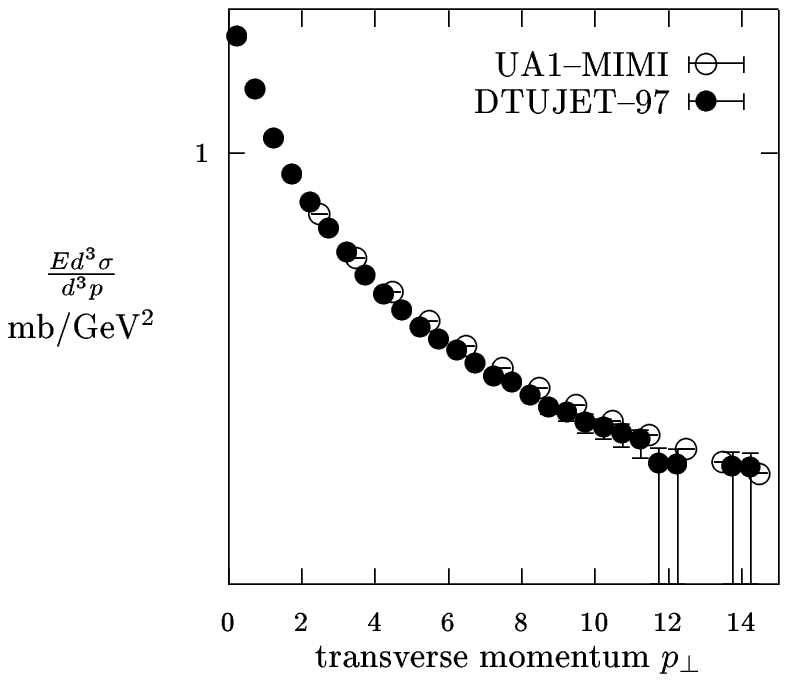,width=8cm,height=7cm}
\end{picture}
\caption{
Comparison of transverse momentum cross sections according to
DPMJET--II.3 at
$\sqrt e$ = 0.63  TeV with
collider data from the UA--1 MIMI Collaboration 
\protect\cite{Bocquet96a}. The experimental date are represented
by the parametrization given by the Experiment.
\label{etjet1800b}
}
\end{figure}

At higher energies (and in non-single
diffractive $p\bar p$ collisions) there are pseudorapidity
distributions from the UA--5 Collaboration
 \protect\cite{Alner86b}
and from the CDF Collaboration \protect\cite{Abe90}. 
In  
 Fig.\ref{prapm22225918nd}
 a very good agreement is found of DMJET--II.3 with these data.
 Still
very often  there is and 
was always (see Fig.\ref{prapm22225918nd})
a disagreement of the models with the UA--5 data 
at the highest pseudorapidity
values. 
 The models predict systematically more
particles at the largest pseudorapidities of the experiment.
This disagreement (if the data would be correct) would  of course be
of importance, if one is
interested in Cosmic Ray cascades, where the particle production
in the fragmentation region is of main interest. Fortunately, a
new independent measurement of the pseudorapidity distribution
in the collider energy range became available recently \cite{Harr97}.
In Fig. \ref{prapm222630nd} the comparison with this new data is
presented and we find a remarkable agreement with DPMJET--II.3
in the large pseudorapidity region.
In Fig.\ref{knapp2} we present the comparison 
(from Ref.\cite{Kcodcomp}) of multiplicity distributions according
to 5 models with the
data from the UA--5 Collaboration 
 \cite{Alner86b}.  
Most of the models describe
at least the high multiplicity tail of the data reasonably well,
however the multiplicity distribution occording to the SIBYLL
model is everywhere rather far from the data.
 We turn to collisions with nuclei. In Fig. \ref{rappar} the
 comparison of DPMJET--II.3 
 is with the rapidity distribution of charged hadrons
 in p-Ar collisions at 200 GeV.  
  In Fig. \ref{rappar}  we compare with the rapidity
  distribution of negatively charged hadrons in central S--S and
  S--Ag collisions.

At least in models with a minijet component we expect good
agreement with data on transverse momentum distributions.
 In Fig.\ref{etjet1800} we compare hadron 
 jet production in DPMJET--II.3
 with 
 data from the CDF--Collaboration \protect\cite{Abe96a}. The jets
 from the model are found out of the Monte Carlo events using a
 jet finding algorithm with the same parameters like the one used by the
 experiment. With a minimum bias Monte Carlo event generator it
 is of course not possible to obtain good statistics on the total
  transverse energy range of the experiment. We find good
  agreement of the jets in the model with the data up to
  $E_{\perp}$ = 30 GeV/c.
The transverse momentum distribution  in a large
$p_{\perp}$ region was  determined by the UA--1--MIMI Collaboration
\cite{Bocquet96a}. In Fig.\ref{etjet1800b} we compare  DPMJET--II.3
results with the parametrization of the data given by this
experiment and we find a good agreement.


\begin{figure}[thb] \centering
\hspace*{0.25cm}
\noindent
\unitlength1mm
\begin{picture}(160,70)
\epsfig{file=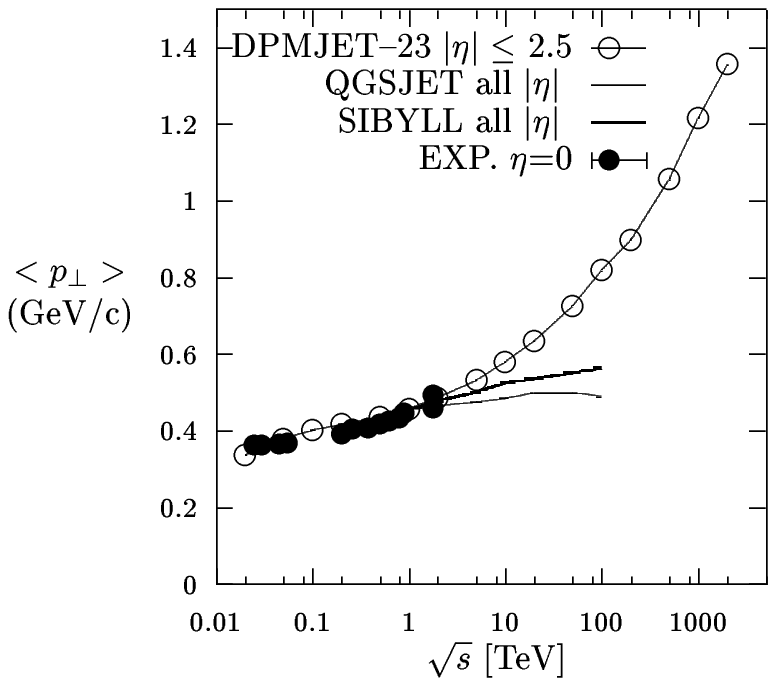,width=8cm,height=7cm}
\end{picture}
\caption{
Average transverse momenta of charged secondaries
produced in $p\bar p$ and $pp$ collisions  calculated from
DPMJET, QGSJET and SIBYLL (The latter two as given in
Ref.\cite{Kcodcomp})
 as function of the center of
mass energy $\sqrt s$ compared to date collected
by the UA--1 Collaboration \protect\cite{Albajar90}.
\label{dpm222ptav}
}
\end{figure}

In Fig.\ref{dpm222ptav} we compare average transverse momenta as
obtained from DPMJET--II.3, QGSJET and SIBYLL
as function of the cms energy $\sqrt
s$ with data collected by the UA--1 Collaboration. This plot
gives at the same time the DPMJET predictions for the average
transverse momenta up to $\sqrt s$ = 2000 TeV and the
predictions of the two other models up to $\sqrt s$ = 100 TeV. 
While at energies where data exist all models agree rather well
with each other and with the data, we find completely different
extrapolations to higher energies. We should note, this are just
the three models with a minijet component. But it seems, that
in spite of the minijets the average transverse momentum in
QGSJET becomes constant at high energies, while it continues to
rise in DPMJET. For me the rise of the average transverse
momentum in DPMJET is connected with the fact, that with the new
parton structure functions since the HERA measurements really
the minijets dominate very much all of hadron production at high
energy. 
We can conclude, there are very big differences in implementing
the minijet components in the models.

 \begin{figure}[thb] \centering
 \hspace*{0.25cm}
 \noindent
 \unitlength1mm
 \begin{picture}(160,70)
\epsfig{file=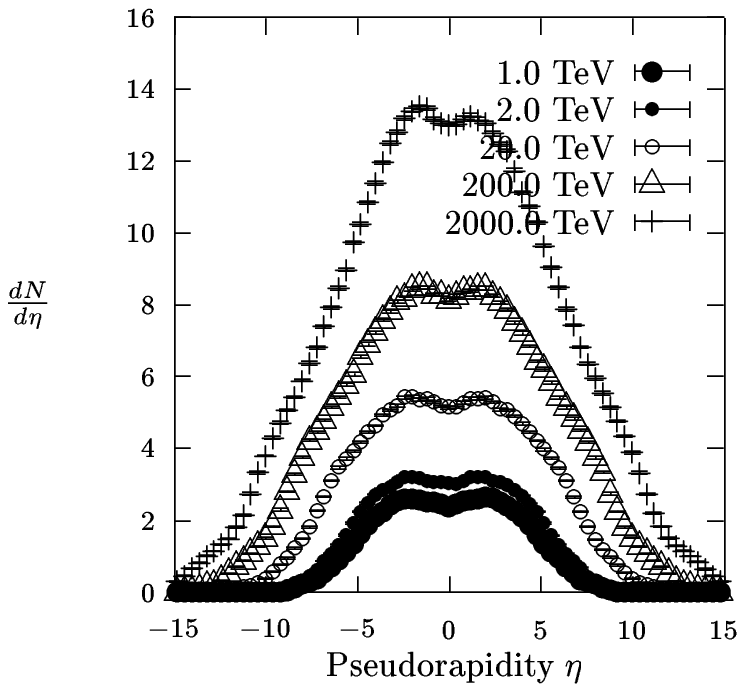,width=8cm,height=7cm}
 \end{picture}
 \caption{
 The development of the pseudorapidity distribution of
 charged hadrons produced in inelastic $pp$ collisions in the  
  in the center of mass energy range between $\sqrt
 s$=1 TeV and $\sqrt s$ = 2000 TeV.
 \label{prapm222}
 }
 \end{figure}

\begin{figure}[thb] \centering
\hspace*{0.25cm}
\noindent
\unitlength1mm
\begin{picture}(160,70)
\epsfig{file=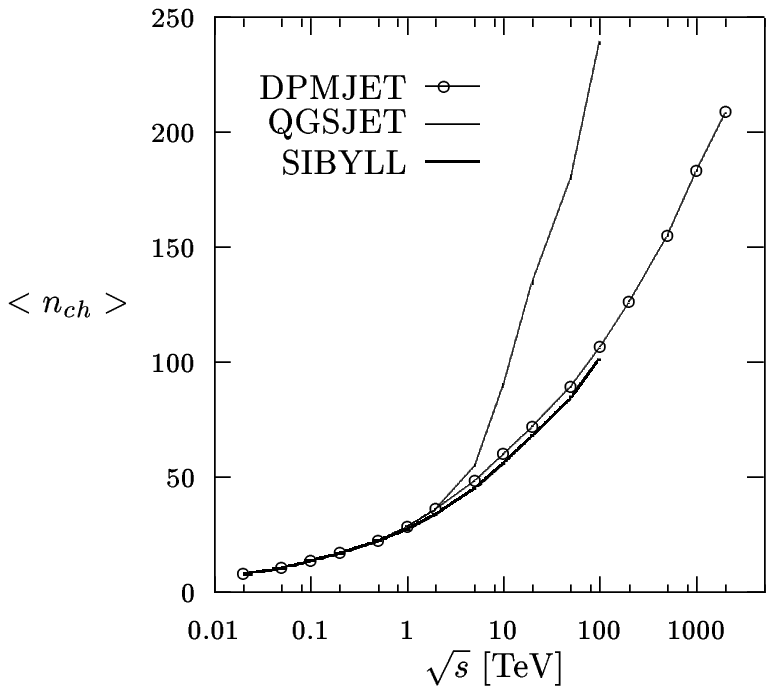,width=8cm,height=7cm}
\end{picture}
\caption{
Rise of the charged multiplicity in inelastic $pp$
collisions according to DPMJET--II.3 
in the center of mass energy range between $\sqrt
s$=0.02 TeV and $\sqrt s$ = 2000 TeV.
At energies between 1 and 100 TeV we plot also the average
multiplicities according to SIBYLL and QGSJET as given in
Ref.\cite{Kcodcomp}.
\label{prapm222b}
}
\end{figure}


\section{Properties of the models in the highest energy region}
 In Fig.\ref{prapm222} the pseudorapidity distributions for
 charged hadrons according to DPMJET--II.3 
 are presented for energies between $\sqrt s$ = 1
 TeV and 2000 TeV. The width of the distributions increases like
 the logarithm of the energy and also the maximum of the curves
 rises like the logarithm of the energy. If we call  the central
 region
 around the two maxima the plateau, then we find the width of 
 this plateau hardly to change with energy.
Fig.\ref{prapm222b}
presents the rise of the total charged  multiplicity with the
cms energy $\sqrt s$ according to DPMJET, QGSJET and SIBYLL. we
find again, at low energies, where data are available, the
models agree rather well. DPMJET and SIBYLL agree in all the
energy range shown. However, QGSJET above the energy of the
TEVATRON extrapolates to higher energies in a completely
different way.
\begin{figure}[thb] \centering
\hspace*{0.25cm}
\noindent
\unitlength1mm
\begin{picture}(160,70)
\epsfig{file=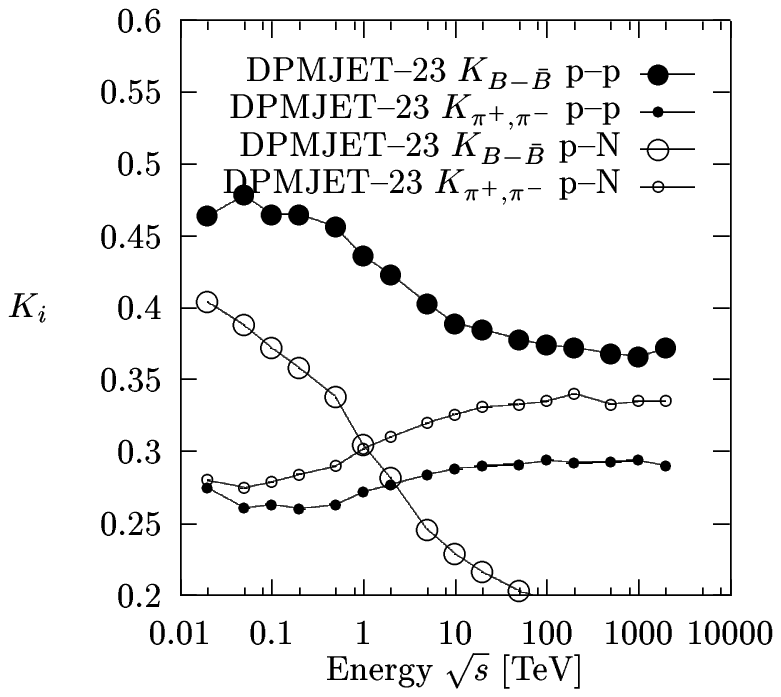,width=8cm,height=7cm}
\end{picture}
\caption{
Laboratory energy fractions  for $B-\bar B$ and
pion  production in $pp$
and $p$--Air collisions according to DPMJET--II.3 
as function of the (nucleon--nucleon)
cms energy  $\sqrt s$.
\label{ptinvm222}
}
\end{figure}

\begin{figure}[thb] \centering
\hspace*{0.25cm}
\noindent
\unitlength1mm
\begin{picture}(160,70)
\epsfig{file=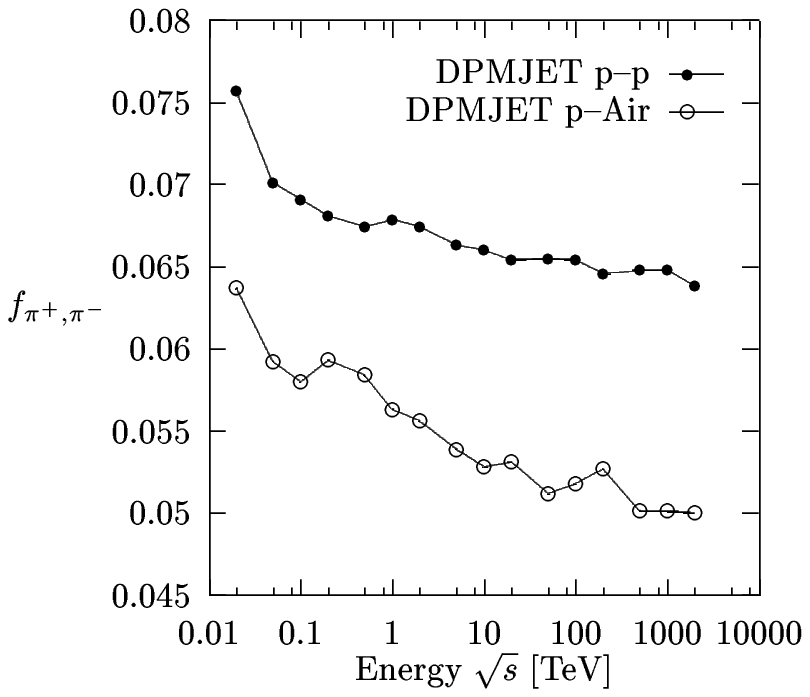,width=8cm,height=7cm}
\end{picture}
\caption{
Spectrum weighted moments for pion production in $pp$
and $p$--Air collisions as function of the (nucleon--nucleon)
cms energy  $\sqrt s$.
\label{dpm222fpi}
}
\end{figure}
In Fig.\ref{ptinvm222} we present  for $pp$ anf p--Air
collisions the energy fractions K  for $B-\bar B$ (baryon -
antibaryon)  and charged pion production.
The cosmic ray spectrum--weighted moments in p--A collisions 
are defined as moments of the $F(x_{lab})$ :
\begin{equation}
Z^{p-A}_i = \int^{1}_0 (x_{lab})^{\gamma -1}
F^{p-A}_i(x_{lab})dx_{lab}
\end{equation}
Here $-\gamma \simeq$ --1.7 is the power of the integral cosmic
ray energy spectrum and $A$ represents both the target nucleus name
and its mass number.
 In Fig.\ref{dpm222fpi} we present the spectrum weighted moments
 for pion production in $pp$ and p--Air collisions as function of
 the cms energy $\sqrt s$ per nucleon. 
We find   
all average values characterizing hadron production:
the cross sections (Fig.\ref{xs222626}), the average transverse
momenta (Fig.\ref{dpm222ptav}) the charged multiplicities 
(Fig.\ref{prapm222b}), and the moments 
 in Figs. \ref{dpm222fpi}
  and
in Fig. \ref{ptinvm222} to change smoothly with
energy in most cases just like the logarithm of the energy.

\section*{Comparison of the models after simulating the Cosmic
Ray cascade}
\begin{figure}
 \begin {center}
 \epsfig{file=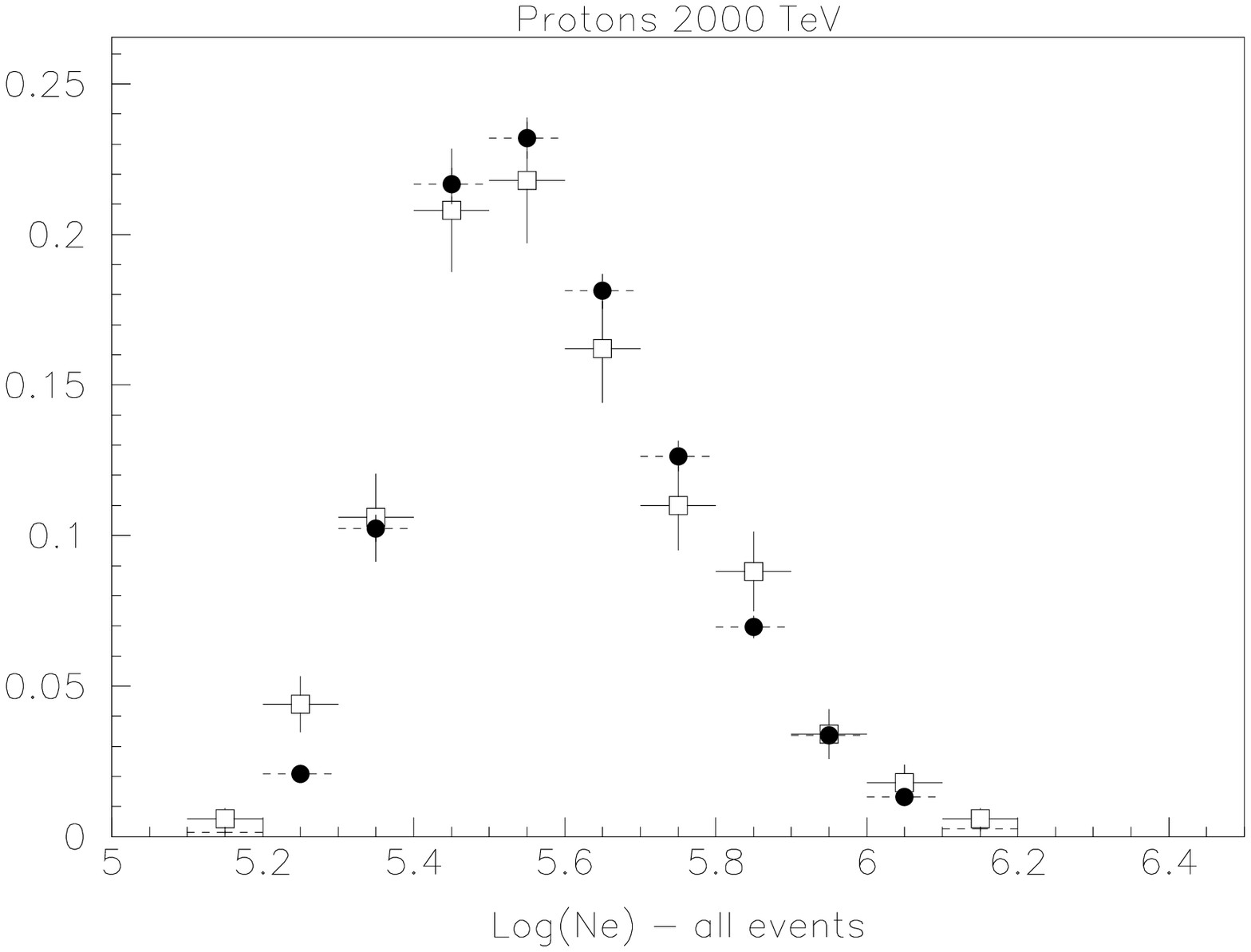,width=7cm,height=7cm}
 \end {center}
 \vskip -2cm  
 \caption{The distribution of the electron number at the
 sampling level
 calculated for 2000 TeV primary protons. 
 }
 \label{p20002}
 \end{figure}

\begin{figure}
 \begin {center}
 \epsfig{file=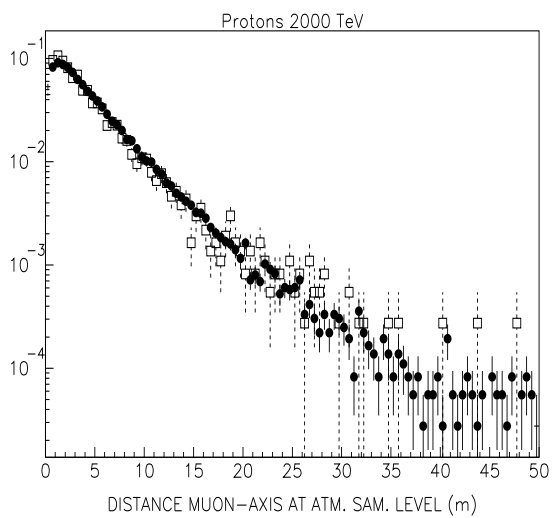,width=7cm,height=7cm}
 \end {center}
 \vskip -2cm  
 \caption{The distance muon-shower axis for E $>$ 1 TeV muons
 calculated for 2000 TeV primary protons. This distribution is
 mainly related to the transverse momentum distribution of pions
 and Kaons in the hadronic collisions.}
 \label{p20004}
 \end{figure}

\begin{figure}
\begin {center}
\epsfig{file=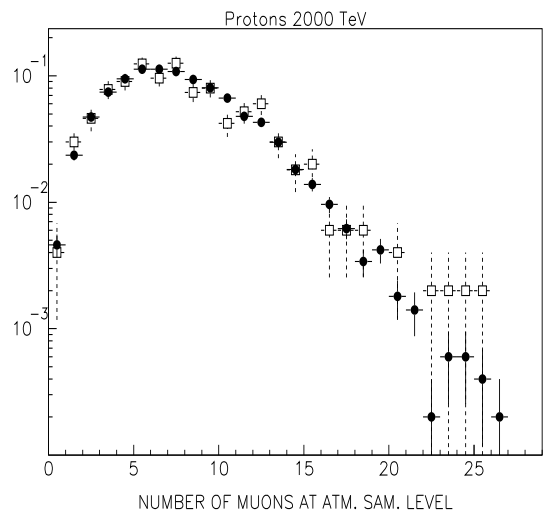,width=7cm,height=7cm}
\end {center}
\vskip -2cm  
\caption{The multiplicity distribution of Muons at the sampling
level}
\label{p20006}
\end{figure}

\begin{figure}
\begin {center}
\epsfig{file=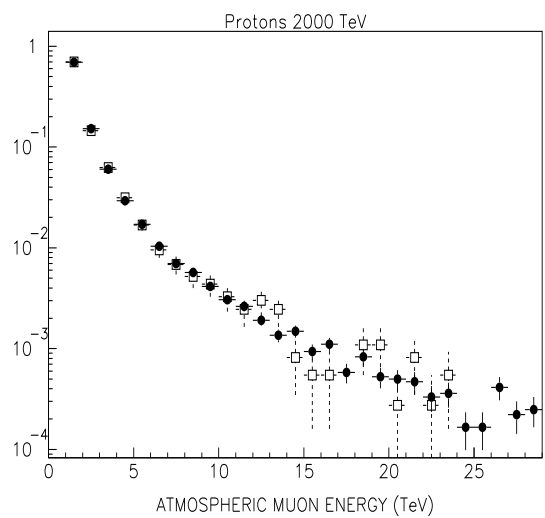,width=7cm,height=7cm}
\end {center}
\vskip -2cm  
\caption{The energy spectrum of E $>$ 1 TeV muons at the sampling
level.}
\label{p20007}
\end{figure}

\begin{figure}
\begin {center}
\epsfig{file=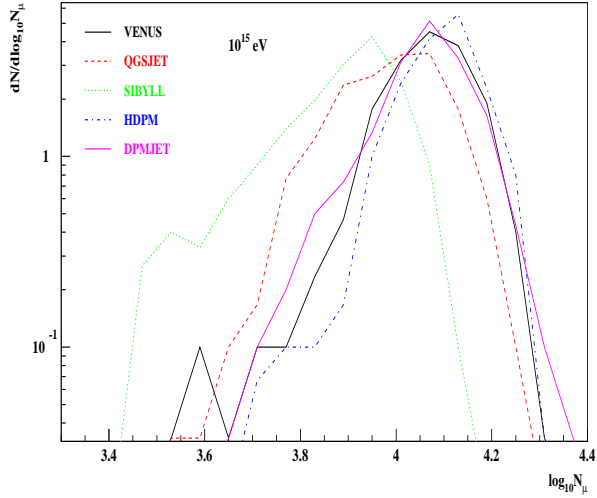,width=8cm,height=7cm}
\end {center}
\vskip -1cm  
\caption{Muon number distribution at ground level 
from proton induced showers with $E = 10^{15}$ eV.
The
distributions were calculated using the CORSIKA shower code
\protect\cite{Kcodcomp} with 5 different event generators for the
hadronic interactions.
}
\label{knapp1}
\end{figure}

 First we present results of a comparison between the cascade
 code HEMAS \cite{HEMAS} using DPMJET as event generator and the
 cascade code CORSIKA \cite{Corsika} using VENUS as event
 generator \cite{Forticoco}. This comparison has been done for
 quantities of interest for the EAS--Top and MACRO experiments
 in the Gran Sasso Lab. The zenith angle is fixed
  at 31 degrees (MACRO/EAS-TOP coincidence direction). The e.m.
  shower size
   and muons above 1 TeV are sampled at 2000 meters a.s.l. (946
   g/cm2 slant
    depth, 810 g/cm2 vert. depth).  The calculations were done
    for primary protons, He nuclei and Fe nuclei with energies
    between 3 and 2000 TeV. Calculated are for each primary
    energy and particle (i) the e.m. shower profile, (ii)the
    Log(e.m. size) at EAS-TOP sampling depth (946 g/cm2),
    (iii)the distance muon-shower axis for E $>$ 1 TeV muons, (iv)
    the muon decoherence for E $>$ 1 TeV muons, (v) the number of
    muons per shower and (vi) the energy spectrum of E $>$ 1 TeV
    muons.
    In Figs. \ref{p20002} to \ref{p20007} we present two of
    these comparisons. A satisfactory agreement is found in
    these plots as well as in all other comparisons at different
    energies and with the other primary particles.
    Next we present two comparisons from the Karls\-ruhe code
    comparison \cite{Kcodcomp}.
    The distributions choosen in this comparison are motivated
    by the interest of the KASKADE\cite{Kaskade} 
    experiment in Karls\-ruhe. 
    In Fig.\ref{knapp1} the Muon
    multiplicity distribution at ground level is calculated for
    primary protons of $E = 10^{15}$ eV. The calculation is done
    with the CORSIKA cascade code using 5 different event
    generators for the hadronic interactions. While again VENUS
    and DPMJET give distributions , which agree very well, it is
    found, that SIBYLL gives a very different distribution
    centered at smaller Muon number. 

\begin{figure}
 \begin {center}
 \epsfig{file=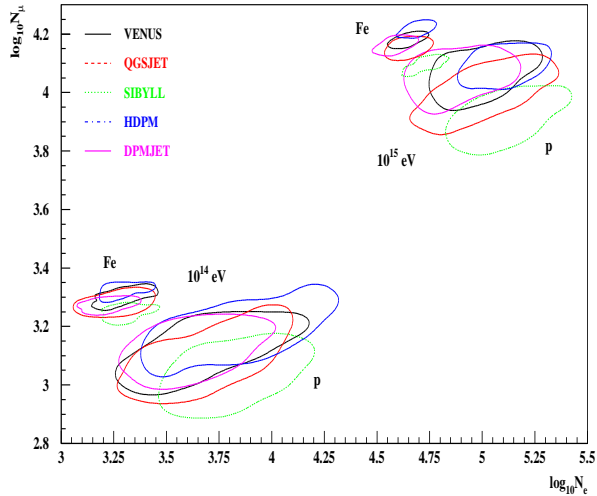,width=8cm,height=7cm}
 \end {center}
 \vskip -1cm  
 \caption{Contours in the $\log_{10}N_{\mu}$ --$\log_{10}N_{e}$
 plane for p and Fe induced showers of $E = 10^{14}$ and
 $10^{15}$ eV .
 }
 \label{knapp3}
 \end{figure}

\begin{figure}
 \begin {center}
 \epsfig{file=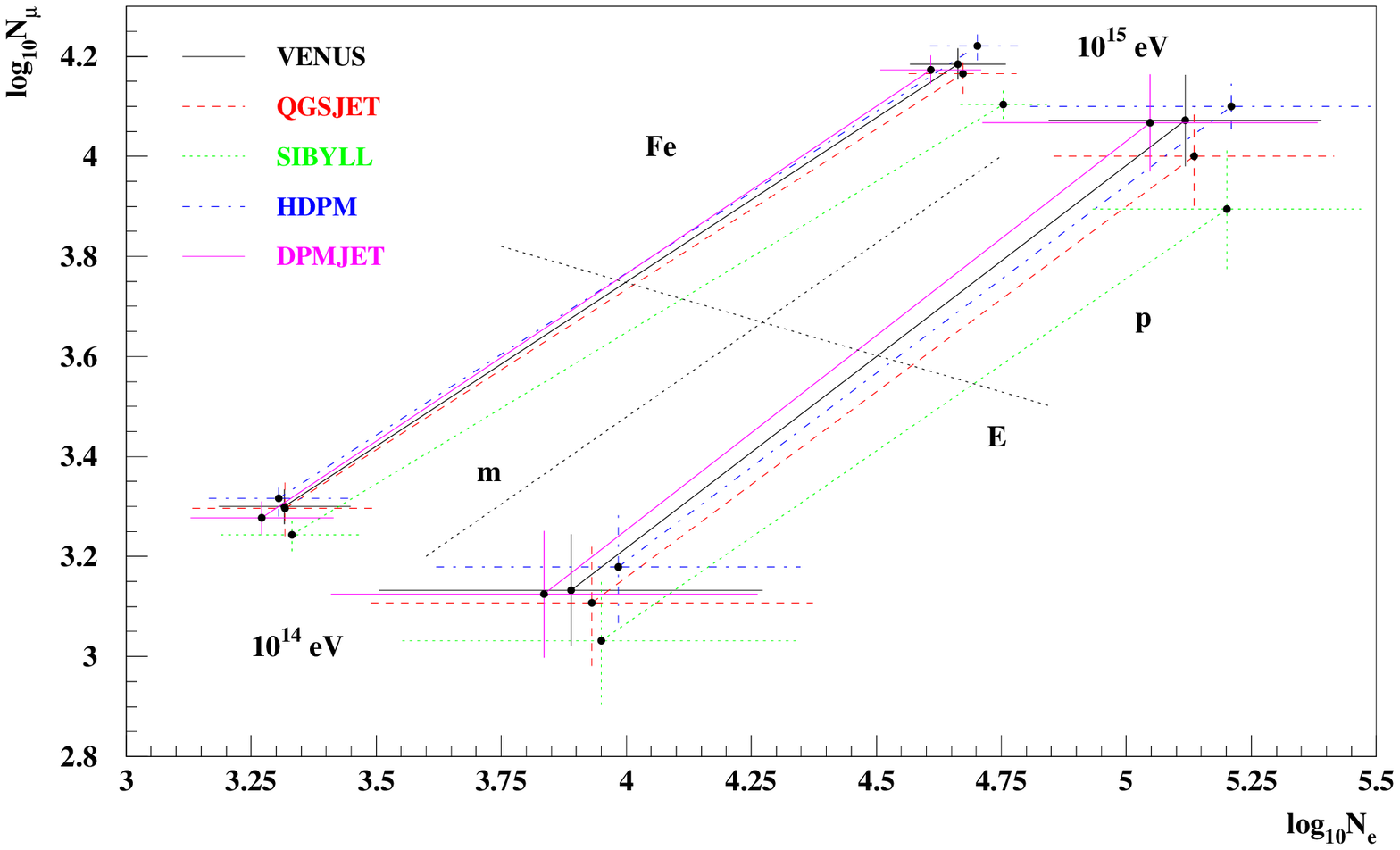,width=8cm,height=7cm}
 \end {center}
 \vskip -1cm  
 \caption{ $\log_{10}N_{\mu}$ over $\log_{10}N_{e}$
 for p and Fe induced showers of $E = 10^{14}$ and
 $10^{15}$ eV .
 Projecting along the lines (m) and (E) one can estimate the
 energy and mass of the primary.
 }
 \label{knapp4}
 \end{figure}

 In Figs. \ref{knapp3} and \ref{knapp4}
 (
The
distributions were calculated using the CORSIKA shower code
\protect\cite{Kcodcomp} with 5 different event generators for the
hadronic interactions.
 )
 Fe and p induced showers with energies of
 $E = 10^{14}$ and $10^{15}$ eV are plotted in the $\log_{10}
 N_{\mu}$ --$\log_{10}N_{e}$ plane (Muon--number --Electron--number
 plane). The distribution of events according to each of the 5
 interaction models for each energy and primary perticle 
 is indicated by contours.
Considering  these plots  calculated
with only one of the models, where Muon number is plotted over
electron number, the impression is, that a
simultaneous measurement of Muon--number and Electron number
allows to determine the primary energy as well as the
composition of the primary component. In these plots we 
see, that for
instance VENUS and DPMJET agree very well, but the contour
according to SIBYLL for Fe projectiles of $E = 10^{15}$ eV
overlaps the VENUS and DPMJET contours for p projectiles.
 From these differences between the models one can
conclude, that at present the systematic errors of the cascade
calculations (and this are just the differences obtained using
different models) 
prevent to identify safely the composition of the primary
component from such measurements.

\section{Conclusions}
I would like to stress, more efforts are needed to extend the
models used to simulate the hadronic interactions in the C.R.
cascade up to the energies to be explored by the Auger
Experiment.

At least at collider energies, where data are available, these
models should agree among themselves and with the data.
Disagreements to data like the ones seen in Fig.\ref{knapp2}
should be removed as soon as discovered.

A much better understanding is needed how to calculate the
minijet component. Certainly, the parton structure functions
used for calculating the minijet cross sections should
correspond to the HERA measurements at small x. But this is
certainly not the only problem. The differences in the
extrapolation to higher energies 
of quantities like average transverse momenta and
charged multiplicities (see Figs. \ref{dpm222ptav} and \ref{prapm222b} )
in the three models implementing minijets are huge. These
differences indicate, that much effort is needed to get a better
understanding of the minijet component.

Another question, where models disagree is the presence at high
energy of an important soft component of hadron production like
in the models with a supercritical pomeron. In minijet models
all rise of the cross sections and of particle production at
high energy is only due to the minijets.

There are (even at energies, where collider data are available,
see Fig. \ref{knapp1}) large differences between the models
after simulating the C.R. cascade. We have to interpret these
differences as the systematic errors of the cascade simulation.
Such large differences could well prevent the interpretation of
otherwise very interesting Cosmic Ray data. In future, C.R. results
should always be interpreted using simulations with some
different models.

It might be dangerous, that at present many of the popular
models are based on the same theoretical foundations (and yet
might differ very much in their results). To be on the safe
side, it would be usefull to construct  models based also on widely
different theoretical concepts (for instance on the string
fusion model \cite{hffrasc}).

Finally, I would like to stress the need for new measurements of
hadron production especially at the highest possible energies.
In particular in the fragmentation region so important for the
cosmic ray cascade, data (like Feynman--x distibutions) from the
TEVATRON collider would be highly welcome.

{\bf  Acknowledgements}

Thanks are due to J.Knapp,  for providing me with some of the
Figures from the Karlsruhe code comparison and due to C.Forti
for providing me with the Figures from the 2000 TeV code
comparison.

%
\bibliographystyle{zpc}
\bibliography{dpm11}
 
%
%
%
%
%


\end{document}

%% file: tables.tex
\newbox\hdbox%
\newcount\hdrows%
\newcount\multispancount%
\newcount\ncase%
\newcount\ncols
\newcount\nrows%
\newcount\nspan%
\newcount\ntemp%
\newdimen\hdsize%
\newdimen\newhdsize%
\newdimen\parasize%
\newdimen\spreadwidth%
\newdimen\thicksize%
\newdimen\thinsize%
\newdimen\tablewidth%
\newif\ifcentertables%
\newif\ifendsize%
\newif\iffirstrow%
\newif\iftableinfo%
\newtoks\dbt%
\newtoks\hdtks%
\newtoks\savetks%
\newtoks\tableLETtokens%
\newtoks\tabletokens%
\newtoks\widthspec%
\immediate\write15{%
CP SMSG GJMSINK TEXTABLE --> TABLE MACROS V. 851121 JOB = \jobname%
}%
\tableinfotrue%
\catcode`\@=11
\def\tstrut{\vrule height3.1ex depth1.2ex width0pt}%
\def\and{\char`\&}
\def\tablerule{\noalign{\hrule height\thinsize depth0pt}}%
\def\thickrule{\noalign{\hrule height\thicksize depth0pt}}%
\def\ctr#1{\hfil\ #1\hfil}%
\tablewidth=-\maxdimen%
\spreadwidth=-\maxdimen%
\def\tabskipglue{0pt plus 1fil minus 1fil}%
\centertablestrue%
\gdef\ARGS{########}
\gdef\headerARGS{####}
\def\@mpersand{&}
{\catcode`\|=13
\gdef\letbarzero{\let|0}
\gdef\letbartab{\def|{&&}}%
\gdef\letvbbar{\let\vb|}%
}
{\catcode`\&=4
\def\ampskip{&\omit\hfil&}
\catcode`\&=13
\let&0
\xdef\letampskip{\def&{\ampskip}}%
\gdef\letnovbamp{\let\novb&\let\tab&}
}
\def\begintable{
   \begingroup%
   \catcode`\|=13\letbartab\letvbbar%
   \catcode`\&=13\letampskip\letnovbamp%
   \def\multispan##1{
      \omit \mscount##1%
      \multiply\mscount\tw@\advance\mscount\m@ne%
      \loop\ifnum\mscount>\@ne \sp@n\repeat%
   }
   \def\|{%
      &\omit\widevline&%
   }%
   \ruledtable
}
\long\def\ruledtable#1\endtable{%
   \offinterlineskip
   \tabskip 0pt
   \def\widevline{\vrule width\thicksize}
   \def\endrow{\@mpersand\omit\hfil\crnorm\@mpersand}%
   \def\crthick{\@mpersand\crnorm\thickrule\@mpersand}%
   \def\crthickneg##1{\@mpersand\crnorm\thickrule
          \noalign{{\skip0=##1\vskip-\skip0}}\@mpersand}%
   \def\crnorule{\@mpersand\crnorm\@mpersand}%
   \def\crnoruleneg##1{\@mpersand\crnorm
          \noalign{{\skip0=##1\vskip-\skip0}}\@mpersand}%
   \let\nr=\crnorule
   \def\endtable{\@mpersand\crnorm\thickrule}%
   \let\crnorm=\cr
   \edef\cr{\@mpersand\crnorm\tablerule\@mpersand}%
   \def\crneg##1{\@mpersand\crnorm\tablerule
          \noalign{{\skip0=##1\vskip-\skip0}}\@mpersand}%
   \let\ctneg=\crthickneg
   \let\nrneg=\crnoruleneg
   \the\tableLETtokens
   \tabletokens={&#1}
   \countROWS\tabletokens\into\nrows%
   \countCOLS\tabletokens\into\ncols%
   \advance\ncols by -1%
   \divide\ncols by 2%
   \advance\nrows by 1%
   \iftableinfo %
      \immediate\write16{[Nrows=\the\nrows, Ncols=\the\ncols]}%
   \fi%
   \ifcentertables
      \ifhmode \par\fi
      \hbox to \hsize{
      \hss
   \else %
      \hbox{%
   \fi
      \vbox{%
         \makePREAMBLE{\the\ncols}
         \edef\next{\preamble}
         \let\preamble=\next
         \makeTABLE{\preamble}{\tabletokens}
      }
      \ifcentertables \hss}\else }\fi
   \endgroup
   \tablewidth=-\maxdimen
   \spreadwidth=-\maxdimen
}
\def\makeTABLE#1#2{
   {
   \let\ifmath0
   \let\header0
   \let\multispan0
   \ncase=0%
   \ifdim\tablewidth>-\maxdimen \ncase=1\fi%
   \ifdim\spreadwidth>-\maxdimen \ncase=2\fi%
   \relax
   \ifcase\ncase %
      \widthspec={}%
   \or %
      \widthspec=\expandafter{\expandafter t\expandafter o%
                 \the\tablewidth}%
   \else %
      \widthspec=\expandafter{\expandafter s\expandafter p\expandafter r%
                 \expandafter e\expandafter a\expandafter d%
                 \the\spreadwidth}%
   \fi %
   \xdef\next{
      \halign\the\widthspec{%
      #1
      \noalign{\hrule height\thicksize depth0pt}
      \the#2\endtable
      }
   }
   }
   \next
}
\def\makePREAMBLE#1{
   \ncols=#1
   \begingroup
   \let\ARGS=0
   \edef\xtp{\widevline\ARGS\tabskip\tabskipglue%
   &\ctr{\ARGS}\tstrut}
   \advance\ncols by -1
   \loop
      \ifnum\ncols>0 %
      \advance\ncols by -1%
      \edef\xtp{\xtp&\vrule width\thinsize\ARGS&\ctr{\ARGS}}%
   \repeat
   \xdef\preamble{\xtp&\widevline\ARGS\tabskip0pt%
   \crnorm}
   \endgroup
}
\def\countROWS#1\into#2{
   \let\countREGISTER=#2%
   \countREGISTER=0%
   \expandafter\ROWcount\the#1\endcount%
}%
\def\ROWcount{%
   \afterassignment\subROWcount\let\next= %
}%
\def\subROWcount{%
   \ifx\next\endcount %
      \let\next=\relax%
   \else%
      \ncase=0%
      \ifx\next\cr %
         \global\advance\countREGISTER by 1%
         \ncase=0%
      \fi%
      \ifx\next\endrow %
         \global\advance\countREGISTER by 1%
         \ncase=0%
      \fi%
      \ifx\next\crthick %
         \global\advance\countREGISTER by 1%
         \ncase=0%
      \fi%
      \ifx\next\crnorule %
         \global\advance\countREGISTER by 1%
         \ncase=0%
      \fi%
      \ifx\next\crthickneg %
         \global\advance\countREGISTER by 1%
         \ncase=0%
      \fi%
      \ifx\next\crnoruleneg %
         \global\advance\countREGISTER by 1%
         \ncase=0%
      \fi%
      \ifx\next\crneg %
         \global\advance\countREGISTER by 1%
         \ncase=0%
      \fi%
      \ifx\next\header %
         \ncase=1%
      \fi%
      \relax%
      \ifcase\ncase %
         \let\next\ROWcount%
      \or %
         \let\next\argROWskip%
      \else %
      \fi%
   \fi%
   \next%
}
\def\counthdROWS#1\into#2{%
\dvr{10}%
   \let\countREGISTER=#2%
   \countREGISTER=0%
\dvr{11}%
\dvr{13}%
   \expandafter\hdROWcount\the#1\endcount%
\dvr{12}%
}%
\def\hdROWcount{%
   \afterassignment\subhdROWcount\let\next= %
}%
\def\subhdROWcount{%
   \ifx\next\endcount %
      \let\next=\relax%
   \else%
      \ncase=0%
      \ifx\next\cr %
         \global\advance\countREGISTER by 1%
         \ncase=0%
      \fi%
      \ifx\next\endrow %
         \global\advance\countREGISTER by 1%
         \ncase=0%
      \fi%
      \ifx\next\crthick %
         \global\advance\countREGISTER by 1%
         \ncase=0%
      \fi%
      \ifx\next\crnorule %
         \global\advance\countREGISTER by 1%
         \ncase=0%
      \fi%
      \ifx\next\header %
         \ncase=1%
      \fi%
\relax%
      \ifcase\ncase %
         \let\next\hdROWcount%
      \or%
         \let\next\arghdROWskip%
      \else %
      \fi%
   \fi%
   \next%
}%
{\catcode`\|=13\letbartab
\gdef\countCOLS#1\into#2{%
   \let\countREGISTER=#2%
   \global\countREGISTER=0%
   \global\multispancount=0%
   \global\firstrowtrue
   \expandafter\COLcount\the#1\endcount%
   \global\advance\countREGISTER by 3%
   \global\advance\countREGISTER by -\multispancount
}%
\gdef\COLcount{%
   \afterassignment\subCOLcount\let\next= %
}%
{\catcode`\&=13%
\gdef\subCOLcount{%
   \ifx\next\endcount %
      \let\next=\relax%
   \else%
      \ncase=0%
      \iffirstrow
         \ifx\next& %
            \global\advance\countREGISTER by 2%
            \ncase=0%
         \fi%
         \ifx\next\span %
            \global\advance\countREGISTER by 1%
            \ncase=0%
         \fi%
         \ifx\next| %
            \global\advance\countREGISTER by 2%
            \ncase=0%
         \fi
         \ifx\next\|
            \global\advance\countREGISTER by 2%
            \ncase=0%
         \fi
         \ifx\next\multispan
            \ncase=1%
            \global\advance\multispancount by 1%
         \fi
         \ifx\next\header
            \ncase=2%
         \fi
         \ifx\next\cr       \global\firstrowfalse \fi
         \ifx\next\endrow   \global\firstrowfalse \fi
         \ifx\next\crthick  \global\firstrowfalse \fi
         \ifx\next\crnorule \global\firstrowfalse \fi
         \ifx\next\crnoruleneg \global\firstrowfalse \fi
         \ifx\next\crthickneg  \global\firstrowfalse \fi
         \ifx\next\crneg       \global\firstrowfalse \fi
      \fi
\relax
      \ifcase\ncase %
         \let\next\COLcount%
      \or %
         \let\next\spancount%
      \or %
         \let\next\argCOLskip%
      \else %
      \fi %
   \fi%
   \next%
}%
\gdef\argROWskip#1{%
   \let\next\ROWcount \next%
}
\gdef\arghdROWskip#1{%
   \let\next\ROWcount \next%
}
\gdef\argCOLskip#1{%
   \let\next\COLcount \next%
}
}
}
\def\spancount#1{
   \nspan=#1\multiply\nspan by 2\advance\nspan by -1%
   \global\advance \countREGISTER by \nspan
   \let\next\COLcount \next}%
\def\dvr#1{\relax}%
\def\header#1{%
\dvr{1}{\let\cr=\@mpersand%
\hdtks={#1}%
\counthdROWS\hdtks\into\hdrows%
\advance\hdrows by 1%
\ifnum\hdrows=0 \hdrows=1 \fi%
\dvr{5}\makehdPREAMBLE{\the\hdrows}%
\dvr{6}\getHDdimen{#1}%
{\parindent=0pt\hsize=\hdsize{\let\ifmath0%
\xdef\next{\valign{\headerpreamble #1\crnorm}}}\dvr{7}\next\dvr{8}%
}%
}\dvr{2}}
\def\makehdPREAMBLE#1{
\dvr{3}%
\hdrows=#1
{
\let\headerARGS=0%
\let\cr=\crnorm%
\edef\xtp{\vfil\hfil\hbox{\headerARGS}\hfil\vfil}%
\advance\hdrows by -1
\loop
\ifnum\hdrows>0%
\advance\hdrows by -1%
\edef\xtp{\xtp&\vfil\hfil\hbox{\headerARGS}\hfil\vfil}%
\repeat%
\xdef\headerpreamble{\xtp\crcr}%
}
\dvr{4}}
\def\getHDdimen#1{%
\hdsize=0pt%
\getsize#1\cr\end\cr%
}
\def\getsize#1\cr{%
\endsizefalse\savetks={#1}%
\expandafter\lookend\the\savetks\cr%
\relax \ifendsize \let\next\relax \else%
\setbox\hdbox=\hbox{#1}\newhdsize=1.0\wd\hdbox%
\ifdim\newhdsize>\hdsize \hdsize=\newhdsize \fi%
\let\next\getsize \fi%
\next%
}%
\def\lookend{\afterassignment\sublookend\let\looknext= }%
\def\sublookend{\relax%
\ifx\looknext\cr %
\let\looknext\relax \else %
   \relax
   \ifx\looknext\end \global\endsizetrue \fi%
   \let\looknext=\lookend%
    \fi \looknext%
}%
\def\tablelet#1{%
   \tableLETtokens=\expandafter{\the\tableLETtokens #1}%
}%
\catcode`\@=12

%% file: frascrpp.bbl
\begin{thebibliography}{10}

\bibitem{Auger}
{The Auger Collaboration}:
\newblock Pierre Auger project design report,
\newblock {},
\newblock Fermilab report,  1995

\bibitem{Agasa}
N.~Chila and et~al.:
\newblock Nucl.Instr.Meth. A 311 (1992) 338

\bibitem{Flyseye}
D.~e.~a. Bird:
\newblock Proc. 23 nd ICRC (Calgary) 2 (1993) 38

\bibitem{Kaskade}
P.~Doll and et~al.:
\newblock The Karlsruhe cosmic ray project KASKADE,
\newblock KFK 4686,
\newblock Karlsruhe report,  1990

\bibitem{Kcodcomp}
J.~Knapp, D.~Heck  and G.~Schatz:
\newblock Comparison of hadronic interaction models used in Air shower
  simulations and their influence on shower development and observables,
\newblock FZKA 5828,
\newblock Karlsruhe report,  1996

\bibitem{DPMJETII}
{ J.~Ranft}:
\newblock \pr D 51 (1995) 64

\bibitem{Ranftsare95}
J.~Ranft:
\newblock DPMJET--II, a Dual Parton Model event generator for hadron--hadron,
  hadron--nucleus and nucleus--nucleus collisions,
\newblock {Proceedings of the second SARE workshop at CERN, 1995,ed. by
  G.R.Stevenson, CERN/TIS--RP/977--05, p. 144,},
\newblock 1997

\bibitem{Dpmjet23}
J.~Ranft:
\newblock DPMJET version II.3 and II.4,
\newblock INFN/AE--97/45,
\newblock Gran Sasso report,  1997

\bibitem{Gluck95a}
M.~Gl\"uck, E.~Reya  and A.~Vogt:
\newblock Z.\ Phys.\ C67 (1995) 433

\bibitem{Lai97}
CTEQ-Collab.:  H.~L. Lai et~al.:
\newblock Phys.\ Rev.\ D55 (1997) 1280

\bibitem{DPMBFR94}
{G.~Battistoni, C.~Forti and J.~Ranft}:
\newblock Astroparticle Phys. 3 (1995) 157

\bibitem{Macro}
S.~e.~a. Ahlen:
\newblock Nucl. Instr. Meth. A 324 (1993) 337

\bibitem{SIBYLL}
R.~S. Fletcher, T.~K. Gaisser, P.~Lipari  and T.~Stanev:
\newblock Phys.\ Rev.\ D50 (1994) 5710

\bibitem{EHQL}
E.~Eichten and et~al.:
\newblock Rev. Mod. Phys. 56 (1984) 579

\bibitem{Corsika}
J.~Knapp and D.~Heck:
\newblock Extensive Air shower simulation with CORSIKA,
\newblock KFK 5196 B,
\newblock Karlsruhe report,  1993

\bibitem{Werner96}
S.~Ostapchenko, T.~Thouw  and K.~Werner:
\newblock Nucl. Phys. B 52B (1997) 3

\bibitem{QGSJET}
N.~Kalmykov and et~al.:
\newblock Physics of Atomic Nuclei 58 (1995) 1728

\bibitem{HPDM}
J.~Capdeveille:
\newblock J.Phys. G 15 (1989) 909

\bibitem{MOCCA}
A.~Hillas:
\newblock J.Phys.G 8 (1982) 1461;1475

\bibitem{CTK87}
A.~Capella, J.~Tran Thanh~Van  and J.~Kwiecinski:
\newblock \prl 58 (1987) 2015

\bibitem{DTUJETZP91}
F.~W. Bopp, A.~Capella, J.~Ranft  and J.~Tran Thanh~Van:
\newblock \zp C51 (1991) 99

\bibitem{DTUJETPR92}
P.~Aurenche, F.~W. Bopp, A.~Capella, J.~Kwiecinski, M.~Maire, J.~Ranft  and
  J.~Tran Thanh~Van:
\newblock \pr D45 (1992) 92

\bibitem{DTUJET92b}
R.~Engel, F.~W. Bopp, D.~Pertermann  and J.~Ranft:
\newblock \prd D46 (1992) 5192

\bibitem{Hahn90}
K.~Hahn and J.~Ranft:
\newblock Phys.\ Rev.\ D41 (1990) 1463

\bibitem{DTUJET93}
{ F.~W.~Bopp, D.~Pertermann, R.~Engel and J.~Ranft,}:
\newblock \pr D 49 (1994) 3236

\bibitem{MRS92}
A.~D. Martin, R.~G. Roberts  and W.~J. Stirling:
\newblock Phys.\ Rev.\ D47 (1993) 867

\bibitem{Donnachie93}
A.~Donnachie and P.~V. Landshoff:
\newblock Phys.\ Lett.\ B296 (1993) 227

\bibitem{Arnison83}
G.~Arnison et~al.:
\newblock Phys.\ Lett.\ B128 (1983) 336

\bibitem{Bozzo84a}
UA4 Collab.:  M.~Bozzo et~al.:
\newblock Phys.\ Lett.\ B147 (1984) 385

\bibitem{Amos85}
N.~A. Amos et~al.:
\newblock Nucl.\ Phys.\ B262 (1985) 689

\bibitem{Bernard87a}
UA4 Collab.:  D.~Bernard et~al.:
\newblock Phys.\ Lett.\ B198 (1987) 583

\bibitem{Alner86a}
UA5 Collab.:  G.~J. Alner et~al.:
\newblock Z.\ Phys.\ C32 (1986) 153

\bibitem{Amos90a}
E710 Collab.:  N.~A. Amos et~al.:
\newblock Phys.\ Lett.\ B243 (1990) 158

\bibitem{Abe93a}
CDF Collab.:  F.~Abe et~al.:
\newblock FERMILAB-PUB-93/234-E,
\newblock 1993

\bibitem{Abe94b}
CDF Collab.:  F.~Abe et~al.:
\newblock Phys.\ Rev.\ D50 (1994) 5518

\bibitem{Abe94d}
CDF Collab.:  F.~Abe et~al.:
\newblock Phys.\ Rev.\ D50 (1994) 5550

\bibitem{Mielke}
{H.H.Mielke, M.F\"oller, J.Engler and J.Knapp }:
\newblock J.Phys. G 20 (1994) 637

\bibitem{Alner86b}
UA5 Collab.:  G.~J. Alner et~al.:
\newblock Z.\ Phys.\ C33 (1986) 1

\bibitem{Abe90}
CDF Collab.:  F.~Abe et~al.:
\newblock Phys.\ Rev.\ D41 (1990) 2330

\bibitem{Harr97}
R.~Harr and et~al.:
\newblock Pseudorapidity distribution of charged particles in $\bar p p$
  collisions at $\sqrt s$ = 630 GeV,
\newblock Hep--ex/9703002,  1997

\bibitem{Adamus88a}
NA22 Collab.:  M.~Adamus et~al.:
\newblock Z.\ Phys.\ C39 (1988) 311

\bibitem{DeMarzo82}
C.~De~Marzo et~al.:
\newblock Phys.\ Rev.\ D26 (1982) 1019

\bibitem{na35qm93}
{NA35 Collaboration, presented by D.R\"ohrich at the QM93 Conference}:
\newblock \np A 566 (1994) 35c

\bibitem{Abe96a}
CDF Collab.:  F.~Abe et~al.:
\newblock Phys.\ Rev.\ Lett.\ 77 (1996) 438

\bibitem{Bocquet96a}
G.~Bocquet et~al.:
\newblock Phys.\ Lett.\ B366 (1996) 434

\bibitem{Albajar90}
UA1 Collab.:  C.~Albajar et~al.:
\newblock Nucl.\ Phys.\ B335 (1990) 261

\bibitem{HEMAS}
{C.~Forti, H.~Bilokon, B.~d'Ettorre Piazzoli, T.K.~Gaisser, L.~Satta and
  T.~Stanev}:
\newblock Phys. \ Rev. \ D42 (1990) 3668

\bibitem{Forticoco}
M.~Ambrosio, C.~Aramo, G.~Battistoni, C.~Forti  and J.~Ranft:
\newblock {to be published},  1997

\bibitem{hffrasc}
{Ferreiro, E.}:
\newblock {these Proceedings},  1997

\end{thebibliography}
